\documentclass[preprint,12pt]{elsarticle}

\usepackage{amssymb}
\usepackage{amsthm}

\usepackage{lineno}
\biboptions{compress}

\usepackage{amsmath}
\usepackage{color}
\usepackage{subcaption}

\usepackage[colorlinks=false, hidelinks]{hyperref}
\usepackage{float} %for option H when placing figures
\usepackage{booktabs} %for top/mid/bottom rule
\newtheorem{theorem}{Theorem}

\providecommand{\refas}[1]{
\noindent
\hspace{\dimexpr-\fboxrule-\fboxsep\relax}\fbox{%
  \begin{minipage}[t]{\linewidth}
   \textit{Reference as:}\\ #1
  \end{minipage}%
}
}

\usepackage[top=2.5cm,bottom=2.5cm,left=3cm,right=3cm]{geometry}

\journal{Pre-peer reviewed version.}

\begin{document}
\begin{frontmatter}

\title{Blinded sample size re-estimation in three-arm trials with `gold standard' design}

\author[add1]{Tobias M\"utze\corref{cor1}}
\author[add1,add2]{Tim Friede}

\address[add1]{Department of Medical Statistics, University Medical Center G{\"o}ttingen, G{\"o}ttingen, Germany}
\address[add2]{DZHK (German Centre for Cardiovascular Research), partner site  G\"ottingen, G\"ottingen, Germany}

\cortext[cor1]{Correspondence: tobias.muetze@med.uni-goettingen.de}

\begin{abstract}
The sample size of a clinical trial relies on information about nuisance parameters such as the outcome variance.
When no or only limited information is available, it has been proposed to include an internal pilot study in the design of the trial.
Based on the results of the internal pilot study, the initially planned sample size can be adjusted. 
In this paper, we study blinded sample size re-estimation in the `gold standard' design for normally distributed outcomes. 
The `gold standard' design is a three-arm clinical trial design which includes an active and a placebo control in addition to an experimental treatment.
We compare several sample size re-estimation procedures in a simulation study assessing operating characteristics including power and type I error. 
We find that sample size re-estimation based on the popular one-sample variance estimator results in overpowered trials. 
Moreover, sample size re-estimation based on unbiased variance estimators such as the Xing-Ganju variance estimator results in underpowered trials, as it is expected since an overestimation of the variance and thus the sample size is in general required for the re-estimation procedure to eventually meet the target power.
Moreover, we propose an inflation factor for the sample size re-estimation with the Xing-Ganju variance estimator and show that this approach results in adequately powered trials. 
Due to favorable features of Xing-Ganju variance estimator such as unbiasedness and a distribution independent of the group means, the inflation factor does not depend on the nuisance parameter and, therefore, can be calculated prior to a trial.
\end{abstract}


\end{frontmatter}

\refas{M\"utze, T. and Friede, T. (2017) Blinded sample size re-estimation in three-arm trials with `gold standard' design. Statistics in Medicine \textit{(In press)} \url{http://onlinelibrary.wiley.com/doi/10.1002/sim.7356/full}}

% \linenumbers

%\let\thefootnote\relax\footnotetext{This is the pre-peer reviewed version of an article accepted in \textit{Statistics in Medicine}.}

\section{Introduction}
Sample size planning for clinical trials is often accompanied by difficulties arising from uncertainty about the primary outcome. 
The implicated misspecification of nuisance parameters during the sample size planning might result in a severely under- or overpowered trial potentially having a negative effect on the success of the study.
To limit the effect of nuisance parameter misspecification, trial designs can allow for adjusting the initially set sample size while the trial is on-going. 
This adjustment is referred to as sample size re-estimation or sample size re-calculation.
The sample size re-estimation can either be based on treatment effect estimates or on nuisance parameter estimates \cite{Proschan2009}.
In this manuscript, we focus on sample size re-estimation based on nuisance parameter estimates.
Wittes and Brittain \cite{Wittes1990} proposed to perform the re-estimation with data from a so-called internal pilot study which data are subsequently also incorporated into the final statistical analysis. 
The re-estimation itself can be performed with either blinded or unblinded data \cite{Friede2006}.
In general, unblinding has the disadvantage of potentially introducing a bias \cite{Schulz2002} and is also not recommended from a regulatory point of view  \cite{ICHE9, CHMP}.
For a more detailed discussion of sample size re-estimation, we refer the reader to the reviews \cite{Proschan2009, Friede2006}.\\ \indent
In this manuscript, we study blinded sample size re-estimation in three-arm trials in the `gold standard' design with normally distributed data. 
The `gold standard' design includes an experimental treatment, an active control referred to as reference, and a placebo. 
This trial design allows assessing both non-inferiority of the experimental treatment compared to the reference as well as the superiority of the experimental treatment or the reference over placebo. 
These superiorities are often considered to prove assay sensitivity, the ability of a trial to distinguish an effective from a non-effective treatment.
For the `gold standard' design multiple hypotheses and testing strategies have been proposed \cite{Pigeot2003, Koch2004, Rohmel2010, Hida2011}.
and planning and analzing studies in this design received great attention in past years \cite{Kieser2007, Hasler2008, Mielke2008, Munzel2009, Balabdaoui2009, Kombrink2013, Mutze2016, Mutze2016b}.\\ \indent
Blinded sample size re-estimation has already been studied extensively for two-arm trials with normally distributed data \cite{LawrenceGould1992, Kieser2003, Ganju2009, Friede2011, Friede2013, Lu2016}.
For this trial design, the nuisance parameter is the outcome variance which is also referred to as within group variance. 
Particular emphasis was put on approaches of estimating the nuisance parameter blinded.
The best results in terms of meeting the target power and controlling the type I error rate were achieved when the sample size is re-estimated based on the one-sample variance estimator which simply estimates the outcome variance by the sample variance of the blinded data set. 
This approach works particularly well in two-arm trials since the within-group variance, i.e. the nuisance parameter, dominates the between-group variance. 
Thus, the one-sample variance estimator which estimates a linear combination of within-group variance and between-group variance overestimates the within-group variance only moderately.
This overestimation of the nuisance parameter and thus the sample size is in general required for the re-estimation procedure to eventually meet the target power. 
Especially for smaller pilot studies, sample size re-estimation procedures with unbiased nuisance parameter estimation result in underpowered trials \cite{Friede2013}. \\
This article is structured as follows. 
We discuss examples of clinical trials in essential hypertension and in schizophrenia in Section 2. 
In Section 3, we describe the statistical model, the statistical hypotheses, and the sample size planning.
Section 4 starts with a discussion of blinded variance estimators and the resulting sample size re-estimation procedures. 
Then, the performance of the sample size re-estimation procedures is assessed in a Monte Carlo simulation study.
In Section 5, we propose an inflation factor for the sample size re-estimation approach based on the Xing-Ganju variance estimator and assess the operating characteristics of the proposed method in a Monte Carlo simulation study.
The results are discussed in Section 6. 
We conclude with the appendix about the distribution of the blinded variance estimators.
\section{Clinical trial examples}
In this section, we have a closer look at two clinical trials with active and placebo control in hypertension and schizophrenia. 
We briefly provide some background on those diseases and then present results of clinical trials. 
\subsection{A trial in essential hypertension}
A recent position paper of the European Society of Cardiology \cite{jackson2015improving} recommends adaptive clinical trial designs such as designs including sample size re-estimation as one strategy to improve clinical trials for cardiovascular disease.
One of the modifiable risk factors for cardiovascular disease is essential hypertension \cite{carretero2000essential}. 
A clinical trial in essential hypertension which included both an active and a placebo control was published by Krum \textit{et al.} \cite{krum1998effect}. 
This trial included a placebo control, Enalapril as an active control, and several doses of Bosentan as the experimental treatment. 
Table \ref{table:tensionExample} shows the change from baseline in diastolic and systolic blood pressure for the placebo control, the active control Enalapril, and one of the doses of the experimental treatment Bosentan (500 mg). 
For the purpose of illustration, we focus here on one dose of Bosentan ignoring the others.
\begin{table}[h]
\caption{Changes from baseline to after treatment in blood pressure as reported in \cite[Table 2]{krum1998effect}.}
\small
\label{table:tensionExample}
\begin{center}
\begin{tabular}{l*{1} {c}{c}{c}}
\toprule
& \textbf{Placebo} & \textbf{Enalapril} & \textbf{Bosentan 500 mg}  \\ \midrule
N & 45 & 46 & 45 \\
Mean change (SE) in diastolic pressure & -1.8 (1.0) & -5.8 (1.0) & -5.7 (1.0) \\
Mean change (SE) in systolic pressure & -0.9 (1.7) & -9.0 (1.7) & -8.4 (1.7) \\
\bottomrule
\end{tabular}
\end{center}
\end{table}\\
Due to the similarity of the sample sizes and the identical standard errors, it can be concluded that the outcome variance is identical between the groups. 
From Table \ref{table:tensionExample} the standardized effect $(\mu_P-\mu_R)/\sigma$ can be estimated to be approximately 0.6 and 0.71 for the diastolic  and systolic blood pressure, respectively.
These effects will be considered later on when selecting the scenarios for the Monte Carlo simulation study comparing the various sample size re-estimation procedures.
\subsection{A trial in schizophrenia}
Schizophrenia is a mental illness which affects a person's social behaviour \cite{Mueser2004}. 
As a recent meta-analysis highlighted, clinical trials in schizophrenia often include both an active control and a placebo control \cite{Rutherford2014}. 
The Positive and Negative Syndrome Scale (PANSS) is an often considered outcome measure in such trials. 
The PANSS was introduced by Kay \textit{et al.} \cite{Kay1987} who also showed that the distribution of the PANSS for schizophrenics can be approximated by a normal distribution. 
A recent study by Loebel \textit{et al.} \cite{Loebel2013} assessed the safety and efficacy of Lurasidone for treating schizophrenia. 
The study included Quetiapine XR as an active control and a placebo control. 
The results of the LS mean change of PANSS from baseline to week 6 are summarized in Table \ref{table:schizoExample}.
\begin{table}[h]
\caption{PANSS change from baseline to week 6 as reported in \cite[Table 2]{Loebel2013}. Change is measured as LSMC (Least-Squares Mean Change).}
\small 
\label{table:schizoExample}
\begin{center}
\begin{tabular}{l*{1} {c}{c}{c}{c}}
\toprule
& \textbf{Lurasidone 80} & \textbf{ Lurasidone 160} & \textbf{Quetiapine XR 600} & \textbf{Placebo}  \\ \midrule
\textbf{N} & 125 & 121 & 116 & 120 \\
\textbf{LSMC (SE)} & -22.2 (1.8) & -26.5 (1.8) & -27.8 (1.8) & -10.3 (1.8)\\
\bottomrule
\end{tabular}
\end{center}
\end{table}\\
The standard error of the mean change for the different treatment groups is identical and with the group specific sample sizes being similar, the outcome variance can be assumed to be identical.
The standardized effect $(\mu_P-\mu_R)/\sigma$ can be estimated to be approximately 0.9.
\section{Statistical model, hypothesis testing, and sample size planning}
\subsection{Statistical model and global hypothesis}
The observations in the experimental treatment (E), reference (R), and placebo (P) group are modeled as normally distributed random variables 
\begin{align*}
X_{k i} \sim \mathcal{N}\left(\mu_k, \sigma^2\right),\quad i=1,\ldots, n_{k},\quad k=E,R,P.
\end{align*}
The variance $\sigma^2$ is assumed to be identical across the three treatment groups.
Throughout this manuscript, we assume that smaller values of $\mu_k$ are better.
The total sample size is given by $n=n_E + n_R + n_P$.
The ratio of the group sample size and the total sample size is denoted as $w_{k}=n_{k}/n$. 
In this manuscript, we focus on assessing the non-inferiority of the experimental treatment compared to the reference. 
Several approaches for formulating hypotheses in the `gold standard` design and the respective testing strategies have been proposed and discussed \cite{Koch2004, Rohmel2010, Hida2011, Hauschke2005a, Stucke2012}. 
We  consider non-inferiority of the experimental treatment compared to the reference as the statistical testing problem
\begin{align*}
H_{0}^{ER}: \mu_{E} - \mu_{R} \geq \delta_{ER} 
\quad \text{vs.} \quad
H_{1}^{ER}: \mu_{E} - \mu_{R} < \delta_{ER},
\end{align*}
with $\delta_{ER}>0$ the non-inferiority margin.
Non-inferiority is demonstrated if the hypothesis $H_0^{ER}$ can be rejected. 
In addition to the non-inferiority hypothesis, the superiority of the experimental treatment and the reference over placebo, respectively, are assessed by testing the statistical hypotheses
\begin{align*}
&H_{0}^{EP}: \mu_{P} - \mu_{E} \leq \delta_{EP} 
\quad \text{vs.} \quad
H_{1}^{EP}: \mu_{P} - \mu_{E} > \delta_{EP},\\
&H_{0}^{RP}: \mu_{P} - \mu_{R} \leq \delta_{RP} 
\quad \text{vs.} \quad
H_{1}^{RP}: \mu_{P} - \mu_{R} > \delta_{RP}.
\end{align*}
The superiority margins $\delta_{EP}$ and $\delta_{RP}$ must be non-negative.
We consider a trial in the `gold standard' design to be successful if the non-inferiority and both superiorities can be demonstrate simultaneously.
This leads to the intersection-union test problem
\begin{align*}
H_{0}: H_{0}^{ER} \cup H_{0}^{EP} \cup H_{0}^{RP} 
\quad \text{vs.} \quad
H_{1}: H_{1}^{ER} \cap H_{1}^{EP} \cap H_{1}^{RP}.
\end{align*}
It is worth noting that the re-estimation procedures discussed in this manuscript also apply when the global hypothesis is defined as the union of the local hypothesis $H_{0}^{ER}$ and $H_{0}^{EP}$, that is when the superiority of the reference to placebo is not tested, or $H_{0}^{ER}$ and $H_{0}^{RP}$, that is when the superiority of the experimental treatment to placebo is not tested.
The re-estimation procedures do not depend on the formulation of the global hypothesis because the estimation of the nuisance parameter, that is the outcome variance, is performed independently of any hypothesis. 
\subsection{Planning and analyzing three-arm trials}
The global hypothesis $H_{0}$ is tested at a significance level $\alpha$ by testing the three local hypotheses $H_{0}^{ER}$, $H_{0}^{EP}$, and $H_{0}^{RP}$   at the same level $\alpha$, respectively.
In other words, no multiplicity adjustment is required  for testing the global hypothesis $H_0$ \cite{Berger1982}. 
The local hypotheses can each be tested by e.g. one-sided Student's t-tests. 
The power of the test procedure for the global hypothesis can be approximated by a multivariate normal distribution \cite{Stucke2012}, that is
\begin{align}
\label{eqn:PowerPair}
B(n)=\Phi\left(
t_{\alpha, \nu_{ER}} - \frac{\delta_{ER}^{*} - \delta_{ER}}{\sigma \sqrt{\frac{1}{n_E} + \frac{1}{n_R}}},
t_{\alpha, \nu_{RP}} - \frac{\delta_{RP}^{*} + \delta_{RP}}{\sigma \sqrt{\frac{1}{n_R} + \frac{1}{n_P}}},
t_{\alpha, \nu_{EP}} - \frac{\delta_{EP}^{*} + \delta_{EP}}{\sigma \sqrt{\frac{1}{n_E} + \frac{1}{n_P}}}
\right),
\end{align}
with $\delta_{ij}^{*}=\mu_{i} - \mu_{j},\, (i,j)=(E,R), (R,P), (E,P)$, the mean differences in the alternative and $\nu_{ij}=n_i+n_j-2$ the degrees of freedom of the t-distribution. Here, $\Phi(\cdot)$ is the cumulative distribution function of the three-dimensional multivariate normal distribution $\mathcal{N}_{3}(0, \Sigma)$ with the covariance matrix
\begin{align}
\label{eqn:CovMatrix}
\Sigma = 
\begin{pmatrix}
1 & -\frac{1}{\sqrt{\left(1+\frac{n_{R}}{n_{E}}\right)\left(1+\frac{n_{R}}{n_{P}}\right)}} & \frac{1}{\sqrt{\left(1+\frac{n_{E}}{n_{R}}\right)\left(1+\frac{n_{E}}{n_{P}}\right)}} \\
-\frac{1}{\sqrt{\left(1+\frac{n_{R}}{n_{E}}\right)\left(1+\frac{n_{R}}{n_{P}}\right)}} & 1 & \frac{1}{\sqrt{\left(1+\frac{n_{P}}{n_{R}}\right)\left(1+\frac{n_{P}}{n_{E}}\right)}} \\
\frac{1}{\sqrt{\left(1+\frac{n_{E}}{n_{R}}\right)\left(1+\frac{n_{E}}{n_{P}}\right)}} & \frac{1}{\sqrt{\left(1+\frac{n_{P}}{n_{R}}\right)\left(1+\frac{n_{P}}{n_{E}}\right)}} & 1
\end{pmatrix}.
\end{align}
Then, the required sample size to test the global hypothesis $H_0$ with a power of $1-\beta$ for a given alternative $(\delta_{ER}^{*},\delta_{RP}^{*},\delta_{EP}^{*})$ is the smallest natural number $n$ for a given allocation $(w_{E}, w_{R}, w_{P})$ such that the power $B(n)$ is equal to or greater than the target power $1-\beta$, that is
\begin{align}
\label{eqn:SSdefinition}
n = \min\left\{ \tilde{n}\in \mathbb{N}: B(\tilde{n})\geq 1-\beta \right\}.
\end{align}
\section{Re-estimation procedures and their performance}
\label{Sec:SSR}
In this section, sample size re-estimation strategies from two-arm trials with normally distributed endpoints \cite{Friede2013, Xing2005} are transferred to three-arm trials. 
The sample size re-estimation in three-arm trials in the `gold standard' design with normal data is conducted by replacing the variance $\sigma^2$ in the sample size formula \eqref{eqn:PowerPair} by an estimate $\hat{\sigma}^2$ obtained from the results of the internal pilot study.
Therefore, in this section we start by presenting various approaches for estimating the variance $\sigma^2$. 
Particular attention is paid to approaches which maintain the blinding of the internal pilot study. 
The unblinded variance estimator will be included for the sake of comparison.
The performance of the resulting re-estimation procedures are then compared considering the global hypothesis $H_{0}$. 
The performance characteristics of interest are power and type I error rate. 
\subsection{Estimating the nuisance parameter}
\label{Sec:NuisanceEst}
Throughout this section, the sample $Y_1, \ldots, Y_{n_1}$ denotes the blinded observations from the internal pilot study with $n_{1}$ being the total sample size of the internal pilot study.
\begin{description}
\item[Unblinded pooled variance estimator] 
The idea of re-estimating the sample size based on unblinded data using the sample variance goes back to Stein (1945) \cite{stein1945two} and was introduced to clinical trials by Wittes and Brittain (1990) \cite{Wittes1990}.
Here, the nuisance parameter $\sigma^2$ can be estimated unbiased by the pooled group specific sample variances, that is
\begin{align*}
\hat{\sigma}^{2}_{Pool} = \frac{(n_{1,E}-1)\hat{\sigma}^{2}_{E} + (n_{1,R}-1)\hat{\sigma}^{2}_{R} + (n_{1,P}-1)\hat{\sigma}^{2}_{P}}{n_{1}-3}.
\end{align*}
Here, $n_{1,k}$ ($k=E,R,P$) are the group specific sample sizes in the pilot phase. 
Internal pilot study designs using this unblinded variance estimator have been studied extensively by numerous authors \cite{coffey1999exact, denne1999estimating, wittes1999internal, proschan2000improved, friede2001comparison, proschan2005two}.
\item[Blinded one-sample variance estimator]
%Gould and Shih (1992) \cite{LawrenceGould1992}
Kieser and Friede (2003) \cite{Kieser2003} proposed to estimate the nuisance parameter by the sample variance of the blinded sample, i.e. 
\begin{align*}
\hat{\sigma}^{2}_{OS} = \frac{1}{n_1-1}\sum_{i=1}^{n_1}\left(Y_i - \bar{Y}_{\cdot}\right)^2.
\end{align*}
It should be noted, that this estimator is unbiased if and only if the group means are identical but biased otherwise.
\item[Blinded adjusted one-sample variance estimator]
As shown for two-arm trials by Gould and Shih (1992) \cite{LawrenceGould1992}, under the planning alternative, an unbiased estimator for the nuisance parameter is obtained by adjusting the blinded one-sample variance estimator by its bias.
In the setting considered here, the estimator $\hat{\sigma}^{2}_{OS}$ has a bias of 
\begin{align*}
\operatorname{Bias}(\hat{\sigma}^{2}_{OS}, \sigma^2)
 =  \frac{n}{n_{1}-1}\Delta_{PR}^{2} \left(w_{1,E} (\Delta^{*})^2 + w_{1,R} - (w_{1,E} \Delta^{*} + w_{1,R})^2\right)
\end{align*}
with $\Delta_{PR}=\mu_{P}-\mu_{R}$,  $\Delta^{*}=(\mu_{P}-\mu_{E})/(\mu_{P}-\mu_{R})$, and $w_{1,k}=n_{1,k}/n_{1}$.
The calculation of the bias in shown in the appendix. 
An unbiased estimator for the nuisance parameter $\sigma^2$ is given by
\begin{align*}
\hat{\sigma}^{2}_{OSU} = \hat{\sigma}^{2}_{OS} - \operatorname{Bias}(\hat{\sigma}^{2}_{OS}, \sigma^2).
\end{align*}
This estimator is only unbiased when the assumptions about the parameters $\Delta_{PR}$ and $\Delta^{*}$ are correct.
When applied in the context of blinded sample size re-estimation, the parameters $\Delta_{PR}$ and $\Delta^{*}$ are calculated under the alternative hypothesis $H_1$.
\item[Blinded variance estimator by Xing and Ganju \cite{Xing2005} ]
In a randomized block design with balanced blocks of length $m$ and $T_{k}$ the sum of the observations in block $k$,
an unbiased blind estimator for the nuisance parameter $\sigma^2$ is given by
\begin{align*}
\hat{\sigma}^2_{XG} =  \frac{1}{n_1-m}\sum_{k}\left(T_{k} - \bar{T}_{\cdot}\right)^2
\end{align*}
with $\bar{T}_{\cdot}$ denoting the mean of the sums $T_{k}$.
\end{description}
\subsection{Sample size re-estimation}
The sample size of a three-arm trial depends on the assumed effect sizes $\delta^{*}_{ER}, \delta^{*}_{RP}$, and $\delta^{*}_{EP}$, on the non-inferiority and superiority margins $\delta_{ER}$, $\delta_{RP}$, and $\delta_{EP}$, on the sample size allocation $(w_{E}, w_{R}, w_{P})$, and on the variance $\sigma^2$. 
The blinded sample size re-estimation is performed by substituting the variance $\sigma^2$ in Formula \eqref{eqn:PowerPair} by a blind variance estimator and then calculating the sample size as defined in Equation \eqref{eqn:SSdefinition}.
The resulting re-estimated sample size is denoted as $\hat{n}_{reest}$. 
However, the re-estimated sample size is not necessarily the final sample size of the trial.
For instance, the final sample size cannot be smaller than the pilot study size $n_{1}$.
Wittes and Brittain \cite{Wittes1990} proposed that the final sample size $\hat{n}_{final}$ should not be smaller than the initially planned sample size $\hat{n}$, meaning that the sample size re-estimation cannot reduce the initially planned sample size.
Birkett and Day \cite{Birkett1994}, however, suggested to also consider reducing the initially planned sample size in which case the final sample size would be the maximum of the pilot study size $n_1$ and the re-estimated sample size $\hat{n}_{reest}$.
Gould \cite{Gould1992} noted to additionally define an upper limit for the final sample size $\hat{n}_{final}$ since too large sample sizes are often not feasible in practice due to limitation of resources.
In this manuscript, we allow for downsizing the initially planned sample size but we do not introduce an upper limit for the final size size.
Therefore, the final sample size is given by
\begin{align*}
\hat{n}_{final} = \max\left\{n_{1}, \hat{n}_{reest}\right\}.
\end{align*}
\subsection{Power and sample size of the re-estimation procedures}
\begin{figure}[h]
  \centering
  \includegraphics[width=.9\linewidth]{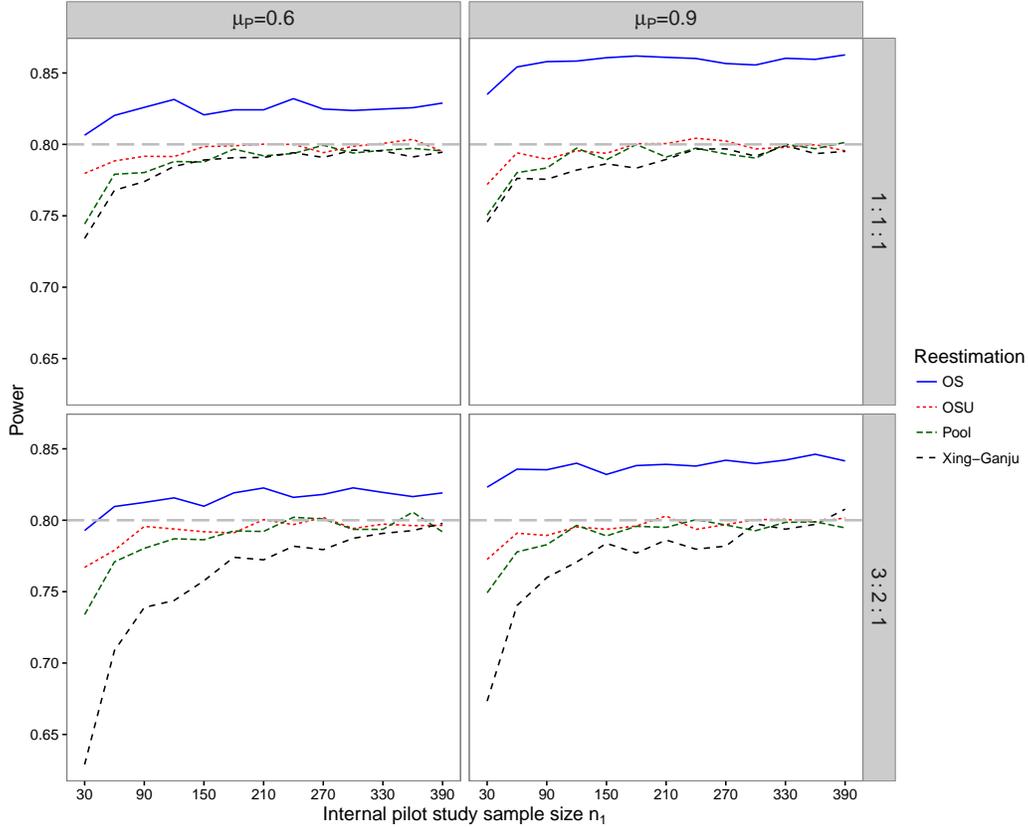}
   \caption{Power of different sample size re-estimation procedures depending on the pilot study size $n_1$. The desired power of $80\%$ is marked as a dashed grey line.}
\label{fig:BSSR_power}
\end{figure}

In this subsection we compare the power and the distribution of the final sample size for the various re-estimation procedures by means of a Monte Carlo simulation study. 
The scenarios of the Monte Carlo simulation study are motivated by the examples in essential hypertension and schizophrenia from Section 2. 
Therefore, we choose the effect sizes $(\mu_P-\mu_R)/\sigma=0.6, 0.9$. 
The parameter combination in the alternative is that the experimental treatment and the reference are equal, that is $\mu_E=\mu_R$.
Table \ref{table:Scenarios} lists the parameters considered in the simulation study.
\begin{table}[h]
\caption{Scenarios for the Monte Carlo simulation study.}
\label{table:Scenarios}
\begin{center}
\begin{tabular}{l*{1} {c}}
\toprule
\textbf{Parameter} & \textbf{Values}  \\ \midrule
One-sided significance level $\alpha$  & $0.025$\\
Target power $1-\beta$ & 0.8\\
Superiority margins $\delta_{EP}, \delta_{RP}$ &  $\delta_{EP} = \delta_{RP} = 0$\\
Non-inferiority margin $\delta_{ER}$ &  $\delta_{ER} = 0.3$\\
Means $\mu_{E}$ and $\mu_{R}$ in the alternative $H_1$ & $\mu_E = \mu_R = 0$\\
Mean $\mu_{P}$ in the alternative $H_1$ & $\mu_{P} = 0.6, 0.9$ \\
Standard deviation $\sigma$ in the alternative $H_1$ & $\sigma = 1$\\
Sample size allocation $n_{E}:n_{R}:n_{P}$  & 1:1:1, 3:2:1\\
Pilot study sample size $n_1$ & $n_1 = 30, 60, \ldots, 390$ \\
\bottomrule
\end{tabular}
\end{center}
\end{table}
For each scenario, we perform $15\,000$ Monte Carlo replications which corresponds to a Monte Carlo error of smaller than 0.0033 for a simulated power of 0.8.
The internal pilot studies are generated with the same allocation as the final trial. 
The Xing-Ganju estimator requires a randomized block design. 
Here, we consider block randomizations with block sizes three and six for allocations 1:1:1 and 3:2:1, respectively.
In Table \ref{table:fixSampleSizes} the samples sizes for a fixed design according to Equation \eqref{eqn:SSdefinition} are listed.
\begin{table}[h]
\caption{Sample sizes for a power of $80\%$ in the fixed design.}
\label{table:fixSampleSizes}
\begin{center}
\begin{tabular}{l*{1} l*{1} {c}}
\toprule
\textbf{Mean} $\mathbf{\mu_{P}}$ & \textbf{Allocation} & \textbf{Sample size $n$}  \\ \midrule
0.6  & 1:1:1 & 525 \\
 & 3:2:1 & 452 \\
%  & 184:184:63 (Optimal) & 431 \\ \hline
0.9 & 1:1:1 & 525 \\
  & 3:2:1 & 438 \\
%  & 178:177:24 (Optimal) & 379 \\
 \bottomrule
\end{tabular}
\end{center}
\end{table}
In Figure \ref{fig:BSSR_power}, the power of the various sample size re-estimation procedures are shown for the scenarios specified in Table \ref{table:Scenarios}.
Figure \ref{fig:BSSR_power} shows that none of the considered sample size re-estimation procedures meets the target power for all considered pilot study sizes.
The sample size re-estimation procedures based on unbiased variance estimators all result in underpowered trials for small pilot studies. 
The one-sample variance estimator results in a sample size re-estimation which considerably overpowers the trials.
The performance of the procedure with the one-sample variance estimator improves as the placebo group mean $\mu_P$ decreases because a decreasing $\mu_{P}$ corresponds to a smaller between-group variability and, thus, to a smaller bias of the variance estimator.
For the same reason, the power of the procedure based on the one-sample variance estimator is closer to the target power when the portion of the sample size allocated to the placebo group decreases. 
Moreover, the power of the re-estimation procedure with the Xing-Ganju estimator decreases as the sample size allocation becomes unbalanced since this corresponds to an increase of the block size from three to six resulting in an higher variability of the estimator.
The sample size re-estimation procedures based on the unbiased one-sample variance estimator and the unblinded pooled variance estimator have a higher power than the re-estimation based on the Xing-Ganju estimator, but also do not meet the target power for small internal pilot studies. 
Additionally, among the sample size re-estimation procedures based on unbiased variance estimators, the Xing-Ganju variance estimator has the slowest convergence against the target power, especially for scenarios with an unbalanced sample size allocation.
However, as mentioned before, using the unbiased one-sample variance estimator and the unblinded pooled variance estimator has the disadvantage of requiring information on the parameters in the alternative and unblinding, respectively.\\
Next we compare the distribution of the final sample size for the sample size re-estimation procedures with the one-sample variance estimator and the Xing-Ganju variance estimator, that are the procedures with the largest and the smallest power, respectively. 
\begin{figure}[h]
  \centering
  \includegraphics[width=.9\linewidth]{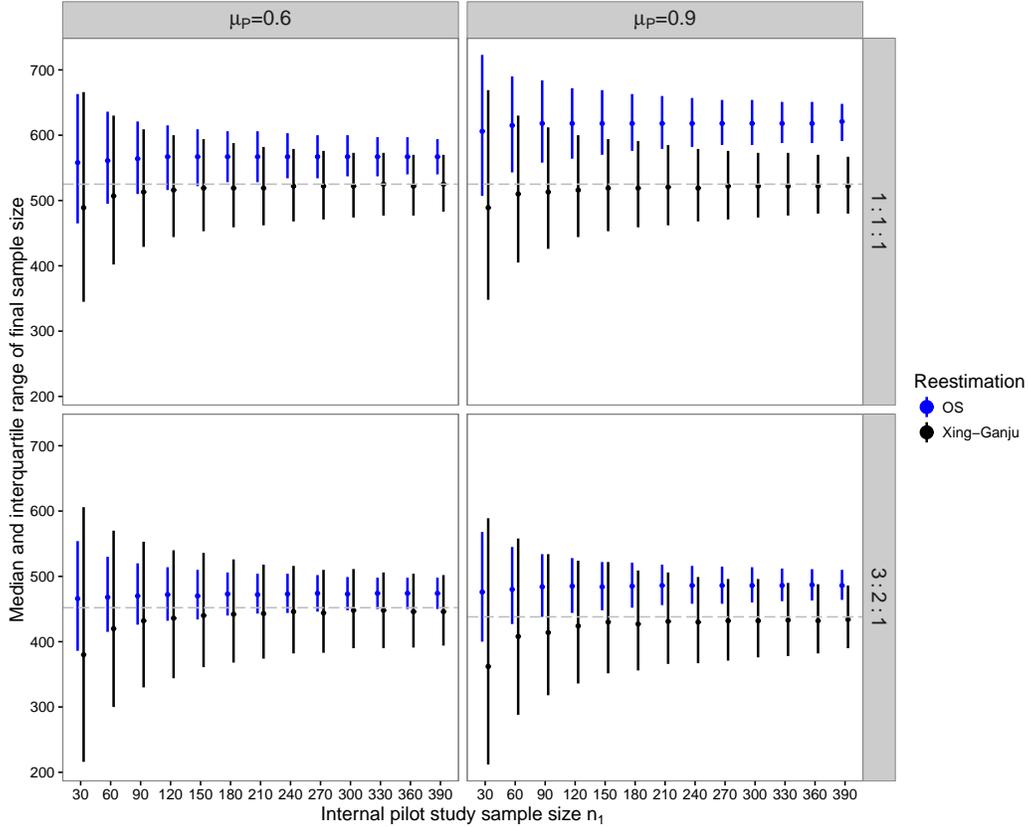}
 \caption{Median and interquartile range of the distribution of the final sample size depending on the pilot study size $n_1$. The dashed grey lines mark the sample size of the fixed design.}
\label{fig:BSSR_n}
\end{figure}\\
Figure \ref{fig:BSSR_n} shows the variability of the final sample size for both procedures decreases as the internal pilot study sample size $n_1$ increases. 
The variability of the final sample size from the procedure based on the Xing-Ganju variance estimator is clearly larger than the variably from the procedure based on the one-sample approach.
These findings are in alignment with what was observed in two-arm trials.
For the re-estimation procedure based on the one-sample variance estimator, the median of the final sample size is larger then the sample size of a fixed design and constant in $n_{1}$.
In contrast, the median of the final sample size of the re-estimation procedure based on the Xing-Ganju variance estimator approaches the sample size of the fixed design  from below as the internal pilot study sample size $n_{1}$ increases.
\subsection{Type I error rate}
\label{Sec:T1Emontecarlo}
\begin{figure}[h]
  \centering
  \includegraphics[width=.9\linewidth]{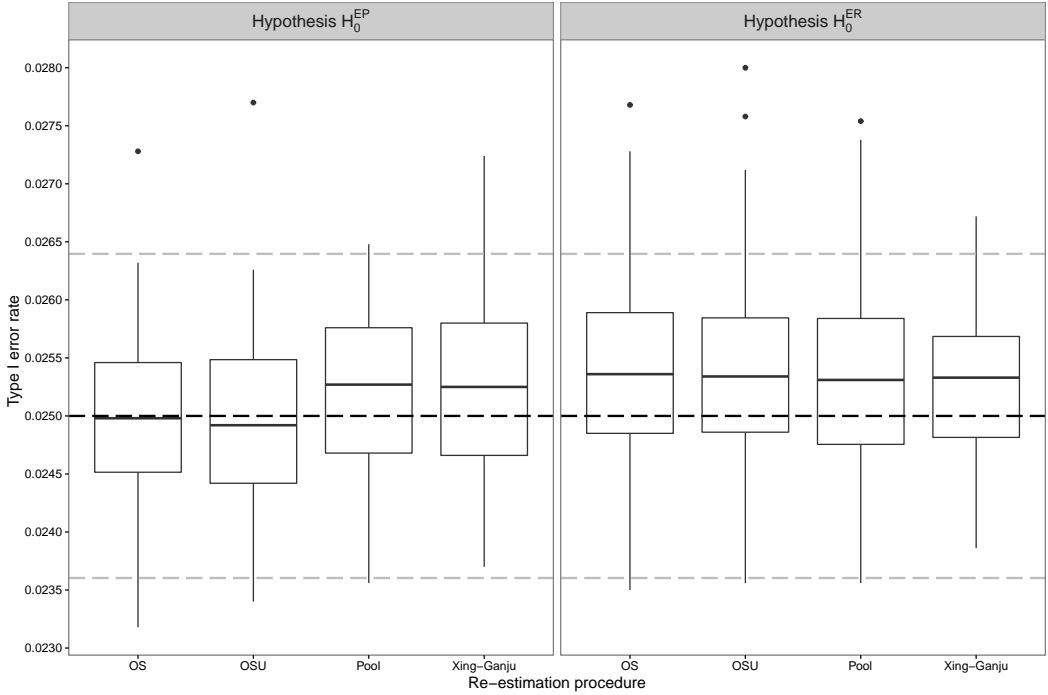}
 \caption{Type I error rates for tests of the hypotheses $H_{0}^{EP}$ and $H_{0}^{ER}$. The dashed black line shows the significance level $\alpha=0.025$ and the dashed grey lines mark the significance level plus/minus two times the Monte Carlo error. Each box plot summarizes the results of 112 scenarios.}
\label{fig:BSSR_typeIerror}
\end{figure}

The type I error rate in designs with a sample size adjustment must be assessed for inflation since a general regulatory requirement for the conduct of adaptive clinical trials is the control of the type I error rate.
The sample size re-estimation is performed for the test of the global hypotheses $H_0$.
However, when testing this hypothesis, the intersection-union test principle is applied rejecting the global hypothesis at significance level $\alpha$ only if all local hypotheses can be rejected at significance level $\alpha$. 
As a consequence, the actual type I error for the global hypothesis is in general not $\alpha$ but smaller. 
Hence, if the sample size re-estimation procedure controls the local type I error rates, control of the global type I error rate is ensured. 
When studying the power, we assumed that superiority of the experimental treatment and the reference over placebo is tested with the margins $\delta_{EP}=\delta_{RP}=0$. 
We study the type I error rate for the same setting. 
However, with $\delta_{ER}>0$, there is not combination of means $(\mu_{E}, \mu_{R}, \mu_{P})$ such that the means are at the boundary of all three local hypotheses $H_{0}^{ER}$, $H_{0}^{EP}$, and $H_{0}^{RP}$.
Therefore, we study the type I error rate separately for the hypothesis $H_{0}^{ER}$ and for the hypotheses $H_{0}^{EP}$, $H_{0}^{RP}$.
The parameters for the simulation study of the type I error rate are listed in Table \ref{table:ScenariosT1E}.
\begin{table}[h]
\caption{Scenarios for the Monte Carlo simulation study of the type I error rate.}
\label{table:ScenariosT1E}
\begin{center}
\begin{tabular}{l*{1} {c}}
\toprule
\textbf{Parameter} & \textbf{Values}  \\ \midrule
One-sided significance level $\alpha$  & $0.025$\\
Target power $1-\beta$ & 0.8\\
Superiority margins $\delta_{EP}, \delta_{RP}$ &  $\delta_{EP} = \delta_{RP} = 0$\\
Non-inferiority margin $\delta_{ER}$ &  $\delta_{ER} = 0.2,\, 0.3,\, 0.4,\, 0.5$\\
Means $\mu_{E}$ and $\mu_{R}$ in the alternative $H_1$ & $\mu_E = \mu_R = 0$\\
Mean $\mu_{P}$ & $\mu_{P} = 0.6,\, 0.9$ \\
Standard deviation $\sigma$ in the alternative $H_1$ & $\sigma = 1$\\
Sample size allocation $n_{E}:n_{R}:n_{P}$  & 1:1:1, 3:2:1\\
Pilot study sample size $n_1$ & $n_1 = 30,\, 90,\, 150,\, \ldots,\, 390$ \\
\bottomrule
\end{tabular}
\end{center}
\end{table}\\
When simulating the type I error rate of the test for the null hypothesis  $H_{0}^{ER}$, the experimental group mean and the reference group mean in the null hypothesis are defined as $\mu_{E,0}=\delta_{ER}$ and $\mu_{R,0}=0$.
In the case of simulating the type I error rate of the test for the null hypotheses $H_{0}^{EP}$ and $H_{0}^{RP}$, the means are chosen as $\mu_{E,0}=\mu_{R,0}=\mu_P$. 
The placebo group mean $\mu_P$ is in both cases defined as listed in Table \ref{table:ScenariosT1E}.
The simulated type I error rates are based on $50\,000$ Monte Carlo replications which corresponds to a Monte Carlo error of 0.0007 for a simulated type I error rate of 0.025.
The simulated type I error rates for the tests of the null hypotheses $H_{0}^{ER}$ and $H_{0}^{EP}$ are summarized as box plots in Figure \ref{fig:BSSR_typeIerror} for the different sample size re-estimation procedures.
The results for the null hypothesis $H_{0}^{RP}$ are not shown here since they are identical to the results for the null hypothesis  $H_{0}^{EP}$.
\\
Figure \ref{fig:BSSR_typeIerror} shows that all sample size re-estimation procedures on average slightly inflate the type I error rate of the test for the non-inferiority hypothesis $H_{0}^{ER}$ by about 0.0005.
The type I error rate of the test for the superiority hypothesis $H_0^{EP}$ is inflated when the sample size re-estimation is conducted with the Xing-Ganju estimator or unblinded with the pooled variance estimator.
The magnitude of the inflation is the same as for the test for $H_{0}^{ER}$. 
The sample size re-estimation procedure with the one-sample variance estimator does not inflate the type I error rate of the test for the superiority hypothesis $H_{0}^{EP}$.
\section{A sample size inflation factor for the Xing-Ganju procedure}
\begin{figure}[h]
  \centering
  \includegraphics[width=0.9\linewidth]{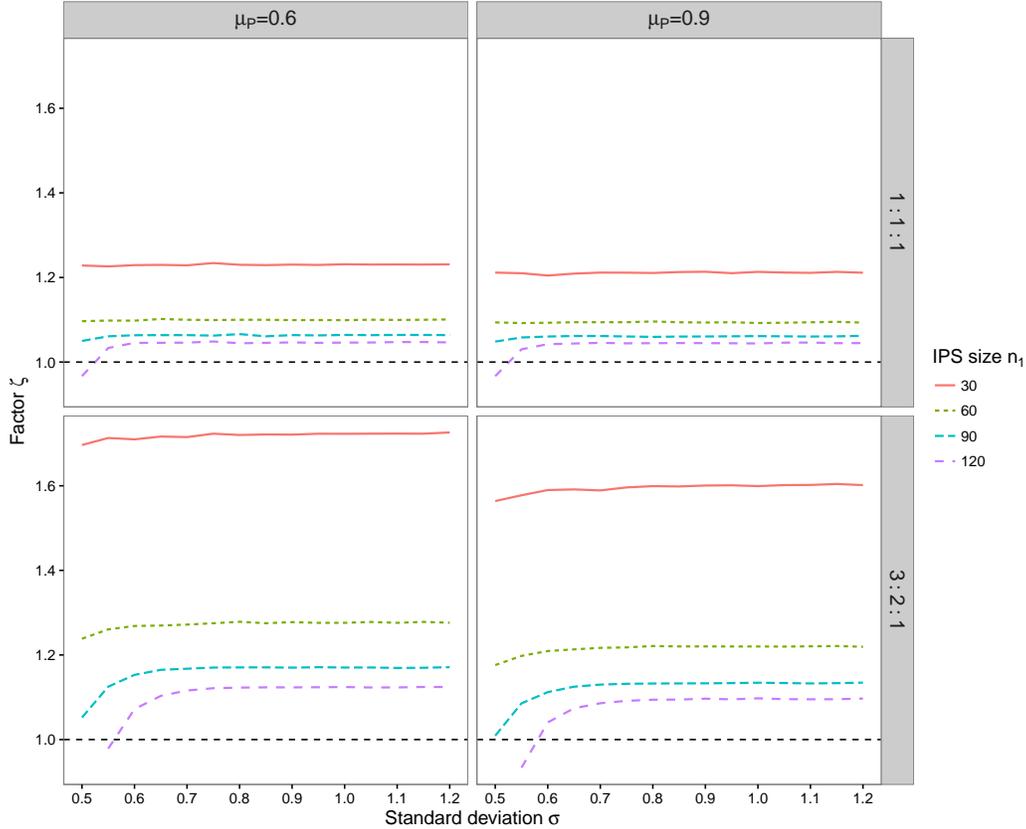}
  \caption{Sample size re-sizing factor $\zeta$ depending on the standard deviation $\sigma$. A reference line at $\zeta=1$ is shown in black.}
\label{fig:BSSRfactor}
\end{figure}

None of the proposed sample size re-estimation procedures meets the desired power for all pilot study sizes in the considered range. 
This motivates introducing a factor to resize the re-estimated sample size such the desired power is met.
Zucker \textit{et al.} \cite{Zucker1999} proposed a resizing coefficient for the re-estimated sample size in a two-arm trial setting.
For the sake of completeness, we present the general idea before applying it to three-arm trials.
Let $B(n)$ be the power function of the testing problem for which the sample size re-estimation is performed, $n(\sigma^2)$ the corresponding sample size as a function of the nuisance parameter,  and $f_{\hat{\sigma}^2}(\cdot)$ the density of the nuisance parameter estimator.
Then, the power of the design with sample size adjustment can be approximated by
\begin{align*}
\text{Power} \approx 
\int_{0}^{\infty} B\left(\tilde{n}\left(x\right)\right)f_{\hat{\sigma}^2}(x) \mathrm{d}x
\end{align*}
with $\tilde{n}\left(x\right)=\max\{n(x), n_1\}$.
This integral only approximates the power of the design with sample size re-estimation because by considering the power function $B(n)$ of the fixed design we ignore that the distribution of the test statistic is not independent of the nuisance parameter estimate from the pilot study.
The resizing coefficient $\zeta\in (0,\infty)$ is the solution to the equation
\begin{align}
\label{Eq:ResizingCoef}
\int_{0}^{\infty} B\left(\tilde{n}_{\zeta}\left(x\right)\right)f_{\hat{\sigma}^2}(x) \mathrm{d}x = 1-\beta .
\end{align}
The final sample size $\tilde{n}_{\zeta}(\cdot)$ is the maximum of the pilot study size and the resized re-estimated sample size, that is 
\begin{align*}
\tilde{n}_{\zeta}\left(x\right)=\max\{\zeta\cdot  n(x), n_1\}.
\end{align*}
In the `gold standard' design, the power function $B(n)$ is given by Equation \eqref{eqn:PowerPair}  and sample size formula $n(\sigma^2)$ by Equation \eqref{eqn:SSdefinition}.
The densities $f_{\hat{\sigma}^2}(\cdot)$ for the nuisance parameter estimators presented in Section \ref{Sec:NuisanceEst} are calculated in Appendix \ref{Sec:DistNuisanceEst}.  
In this section, we focus on the Xing-Ganju variance estimator $\hat{\sigma}^{2}_{XG}$.
We refer to the re-sizing factor as inflation factor since the final sample size from the re-estimation must be inflated. 
As shown in Appendix \ref{Sec:DistNuisanceEst}, the Xing-Ganju variance estimator $\hat{\sigma}^{2}_{XG}$ follows a stretched chi-square distribution with $(b-1)$ degrees of freedom with $b$ the number of blocks in the randomization scheme.
The stretching factor of the stretched chi-square distribution depends on the standard deviation $\sigma$, the pilot study sample size $n_{1}$, and randomized block sizes.
The pilot study size $n_{1}$ and the randomized block sizes are known before conducting the study. 
The standard deviation $\sigma$ is unknown which motivates the use of an internal pilot study in the first place. 
Therefore, the questions remains whether the inflation factor $\zeta$ must be calculated based on the re-estimate standard deviation or if it is constant in the standard deviation $\sigma$ in which case the inflation factor $\zeta$  could be calculate beforehand based on the pilot study size and the block size. 
In the following, we study the effect of the standard deviation $\sigma$ on the re-sizing factor for the scenarios from Table \ref{table:Scenarios}.
In Figure \ref{fig:BSSRfactor} the re-sizing factor $\zeta$ is plotted against $\sigma$ for various pilot study sizes $n_1$.
The size of the randomized blocks for allocations 1:1:1 and 3:2:1 are three and six, respectively.
\\
Figure \ref{fig:BSSRfactor} shows that the factor $\zeta$ is constant in the standard deviation $\sigma$ for small pilot study sizes ($n_1=30, 60$).
For larger pilot study sample sizes ($n_1\geq 90$), the  factor $\zeta$ increases in $\sigma$ first and becomes constant for larger standard deviations $\sigma$. 
For small standard deviations $\sigma$, the factor $\zeta$ decreases as the standard deviation decreases since the smaller the standard deviation the smaller is the difference between required sample size of the fixed design and the pilot study. 
For instance, for the scenario with $\mu_{P}=0.9$,  $\sigma=0.55$, and allocation 3:2:1, the sample size of the fixed design is 134. 
In this case, the factor $\zeta$  for an internal pilot study of size $n_{1}=120$ would depend on the assumed standard deviation. 
Moreover, for some pilot studies sizes, the re-sizing factor cannot be calculated since the pilot study is already larger than the final sample size of the fixed design.
Concluding, the sample size inflation factor $\zeta$ is constant for practical relevant scenarios in which the pilot study sample size is not almost as big as the required sample size of the fixed design. 
Therefore, when we study the power of the sample size re-estimation design with inflated sample size, we calculate the factor based on the pilot study size and the block size for a standard deviation $\sigma$ in the constant range of the inflation factor $\zeta$.
Next, we study the power of the sample size re-estimation design when the re-estimated sample size is adjusted with the factor $\zeta$. 
We again consider the scenarios from Table \ref{table:Scenarios}. 
The results are based on 15\,000 Monte Carlo simulations.
\begin{figure}[h]
  \centering
  \includegraphics[width=0.9\linewidth]{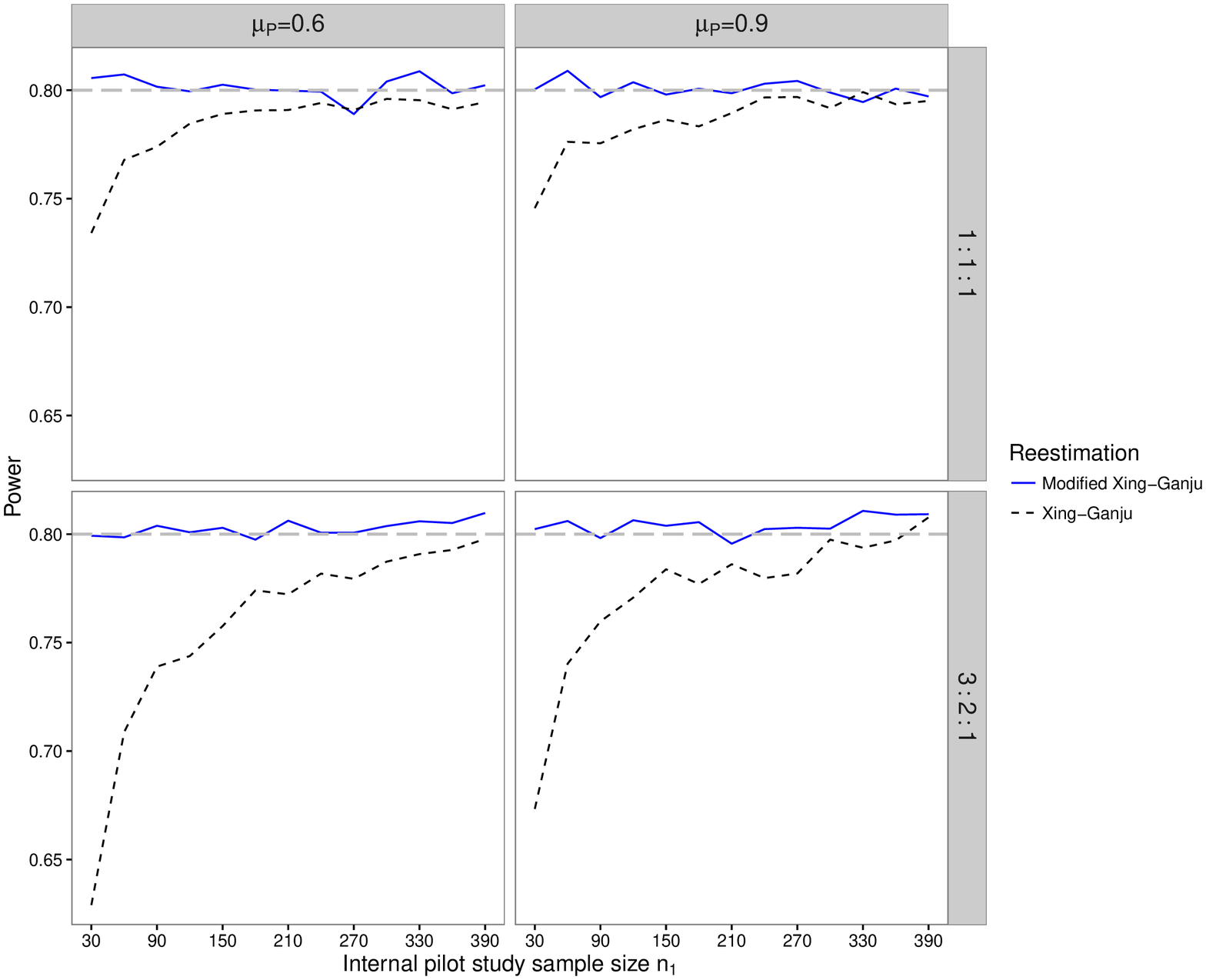}
  \caption{Power of designs with sample size adjustment with and without inflated sample size depending on the internal pilot study sample size $n_{1}$. The desired power of $80\%$ is marked as a dashed grey line.}
\label{fig:BSSRfactorPower}
\end{figure}
Figure \ref{fig:BSSRfactorPower} shows that applying the re-sizing factor to the re-estimated sample size results in a re-estimation procedure which meets the target power for all considered internal pilot study sizes.
Moreover, the sample size re-estimation based on the Xing-Ganju estimator and a re-sized sample size is the only one of the studied re-estimation procedures which meets the desired power for all considered pilot study sample sizes. 
However, inflating the re-estimated sample size with a factor $\zeta$ also increases the variability of the re-estimated sample size. 
As seen in Figure \ref{fig:BSSRfactor}, the inflation factor $\zeta$ increases when the internal pilot study decreases or the block size increases.
The simulation of the type I error rate of the sample size re-estimation procedure with the Xing-Ganju estimator and sample size re-sizing factor is performed analogously to the simulation study in Section \ref{Sec:T1Emontecarlo}.
The type I error rate of the procedures with and without re-sizing factor are similar and will not be shown here.\\
In theory, the presented approach for calculating a sample size re-sizing factor $\zeta$ by solving Equation \eqref{Eq:ResizingCoef} can be applied analogously to the other variance estimators. 
However, in contrast to the Xing-Ganju estimator, the distribution of the one-sample variance estimator does depend on the true means and the factor $\zeta$ depends on the unknown standard deviation $\sigma$. 
Therefore, the factor $\zeta$ cannot be calculated prior to the start of the pilot study without making assumptions on the standard deviation.
\section{Discussion}
In this manuscript, we studied blinded sample size re-estimation for three-arm non-inferiority trials in the `gold stan\-dard' design with non-inferiority defined by an absolute margin.
The re-estimation procedures are based on an estimator of the nuisance parameter obtained from the results of an internal pilot study.
Transferring ideas from sample size re-estimation in two-arm trials, as blinded variance estimators we considered the one-sample variance estimator, a bias adjusted version of the one-sample variance estimator, and the Xing-Ganju estimator which is unbiased and based on randomized block sums. 
The pooled variance estimator from the unblinded data was included for the sake of comparison.
A Monte Carlo simulation study showed that the blinded sample size re-estimation procedure based on the one-sample variance estimator in general results in overpowered trials, except when the between-group variability and the internal pilot study sample size are small.
The sample size re-estimation based on the other variance estimators, which are all unbiased, underpower trials for small internal pilot study sample sizes.
Moreover, we introduced an inflation factor for the sample size re-estimated based on the Xing-Ganju estimator.
The sample size inflation factor is calculated based one the internal pilot study sample size and the block size for a specific alternative and does not depend on the nuisance parameter $\sigma$.
The resulting sample size re-estimation procedure is the only procedure which meets the target power for all considered internal pilot study sizes.
The different sample size re-estimation procedures slightly inflate the type I error rate.
However, procedures for controlling the type I error rate in designs with sample size re-estimation based on an internal pilot study have been suggested  \cite{kieser2000re, Friede2010b, friede2010blinded} and can be applied easily to the sample size re-estimation procedure proposed in this article. 
Concluding, sample size re-estimation in three-arm non-inferiority trials should be performed based on the Xing-Ganju estimator if the data is available in randomized blocks and the re-estimated sample size should be inflated with the factor $\zeta$ as presented in Section 5. 
Alternatively, if overpowering a trial is not an issue and, in particular, if the between-group variability is small, the sample size re-estimation should be conducted based on the one-sample variance estimator. \\ \indent
We studied sample size re-estimation for a non-inferiority hypothesis defined by an absolute margin.
Non-inferiority can also be defined based on the ratio of mean difference, that is $(\mu_E -\mu_P)/(\mu_R-\mu_P)$ \cite{Pigeot2003}.
The resulting hypothesis is in general referred to as the \textit{retention-of-effect} hypothesis.
For the retention-of-effect hypothesis, the outcome variance $\sigma^2$ is still the nuisance parameter and the estimators presented in this manuscript  can be considered to estimate $\sigma^2$ from results of the internal pilot study. 
Therefore, a similar performance of the different sample size re-estimation approaches can be expected in a design with non-inferiority defined by the retention-of-effect hypothesis. \\ \indent
Adjusting a primary outcome for baseline covariates can help minimizing the bias of a statistical analysis and additionally improve the analysis' efficiency \cite{CHMPcovariates}. 
Future research on sample size re-estimation in three-arm trials could focus on designs with baseline covariates. 
For two-arm trials with covariate adjustment, Friede and Kieser \cite{Friede2010b} studied sample size re-estimation based on a one-sample variance estimator and Ganju and Xing \cite{Ganju2009} proposed a sample size re-estimation procedure based on randomized block sums. 
Friede and Kieser \cite{Friede2011} studied sample size re-estimation in internal pilot study designs with ANCOVA based on a one-sample variance estimator.
These approaches could be extended to three-arm trials in the `gold standard' design. 
\appendix
\section{Distributions of the blinded variance estimators}
\label{Sec:DistNuisanceEst}
In this appendix we calculate the distribution of the blinded variance estimators. Similar results for two-arm trials were obtained by Friede and Kieser (2013) \cite{Friede2013}. 
In this appendix we denote the blinded sample by $\mathbf{Y}_{n_{1}}=(Y_{1}, \ldots, Y_{n_{1}})$.
For known group sample sizes $n_{1,k}$, $k=E,R,P$, the vector $\mathbf{Y}_{n_{1}}$ is multivariate normally distributed with mean vector $\boldsymbol{\mu}$ and variance matrix $\sigma^2 \mathbf{I}_{n_{1}}$.
The mean vector $\boldsymbol{\mu}$ is an unknown permutation of the vector with $n_{1,k}$ entries equal to $\mu_k$ for $k=E,R,P$. 
Furthermore, we recall the following theorem which follows from \cite[Theorem 7.3]{rao1972}.
\begin{theorem}
\label{Theorem:DistQuadraticSum}
Let $Y\sim \mathcal{N}_{n}(\boldsymbol{\mu}, \mathbf{V})$ be multivariate normally distributed with $\boldsymbol{\mu}\in \mathbb{R}^n$, $\mathbf{V}\in \mathbb{R}^{n\times n}$, $|\mathbf{AV}|\neq 0$, $\mathbf{A}=\mathbf{A}^{\prime}\in \mathbb{R}^{n\times n}$, and $k=rk(\mathbf{AV})=rk(\textbf{A})$.
Then, $\mathbf{Y^{\prime} A Y} \sim \chi^{2}_k(\lambda)$ with $\lambda=\boldsymbol{\mu}^{\prime} \mathbf{A} \boldsymbol{\mu}$ if and only if $\mathbf{AV}$ is idempotent.
\end{theorem}
\subsection{One-sample variance estimator}
To determine the distribution of the one-sample variance estimator $\hat{\sigma}^2_{OS}$, we define the matrix $\mathbf{A}:=\frac{1}{\sigma^2}\mathbf{P}_{n}$ with $\mathbf{P}_{n}$ the centralizing matrix of dimension $n$.
We note the relation
\begin{align*}
\mathbf{Y^{\prime}A Y} 
= \frac{n_{1}-1}{\sigma^2} \hat{\sigma}^{2}_{OS}.
\end{align*}
With Theorem \ref{Theorem:DistQuadraticSum} it follows that 
\begin{align*}
\mathbf{Y^{\prime}A Y}  \sim \chi^2_{n_{1}-1}(\lambda)
\end{align*}
with 
\begin{align*}
\lambda=\frac{1}{\sigma^2}\sum_{k=E,R,P} n_{k}(\mu_k-\bar{\mu})^2.
\end{align*}
Therefore, the one-sample variance estimator follows a stretched noncentral chi-squared distribution
\begin{align*}
\hat{\sigma}^2_{OS}\sim \frac{\sigma^2}{n_{1}-1} \chi^2_{n_{1}-1}(\lambda).
\end{align*}
The density function $f_{\hat{\sigma}^2_{OS}}$ is given by
\begin{align*}
f_{\hat{\sigma}^2_{OS}}(x) = 
g\left(x\frac{(n_{1}-1)}{\sigma^2}\right) \frac{(n_{1}-1)}{\sigma^2}
\end{align*}
with $g$ the density of a $\chi^2_{n_{1}-1}(\lambda)$-distributed random variable.
\subsection{Xing-Ganju variance estimator}
The variance estimator $\hat{\sigma}^2_{XG}$ is based on the block sums $T_{k}$ with $k=1,\ldots,b=n_1/m$. 
We define the group sample sizes within a block as $m_{k},\, k=E,R,P$, with $m$ the block size. 
Then, the block sums are normally distributed,
\begin{align*}
T_i \sim \mathcal{N}\left(m_{E}\mu_E+m_{R}\mu_R+m_{P}\mu_P, m\sigma^2\right).
\end{align*}
Applying Theorem \ref{Theorem:DistQuadraticSum} gives the distribution of $\hat{\sigma}^2_{XG}$, that is
\begin{align*}
\hat{\sigma}^2_{XG} \sim \frac{m\sigma^2}{n_1-m} \chi^2_{b-1}.
\end{align*}
The density function $f_{\hat{\sigma}^2_{XG}}$ is given by
\begin{align*}
f_{\hat{\sigma}^2_{XG}}(x) = 
g\left(x\frac{(n_{1}-m)}{m\sigma^2}\right) \frac{(n_{1}-m)}{m\sigma^2}
\end{align*}
with $g$ the density of a $\chi^2_{b-1}$-distributed random variable.
 \section*{Acknowledgement}
Tobias M\"utze is supported by the DZHK (German Centre for Cardiovascular Research) under grant GOE SI 2 UMG Information and Data Management.

%% References


\bibliographystyle{model1-num-names}
%\bibliographystyle{abbrvnat}



\end{document}